\documentclass[12pt,preprint]{aastex}
\shorttitle{Search for PeV Gamma Rays}
\shortauthors{Amenomori et al.}

\begin{document}


\title{Observation of PeV Gamma Rays from the Monogem Ring \\with the Tibet Air Shower Array}


\author{
M.~Amenomori,\altaffilmark{1}
S.~Ayabe,\altaffilmark{2}
D.~Chen,\altaffilmark{3}
S.W.~Cui,\altaffilmark{4}
Danzengluobu,\altaffilmark{5}
L.K.~Ding,\altaffilmark{4}
X.H.~Ding,\altaffilmark{5}
C.F.~Feng,\altaffilmark{6}
Z.Y.~Feng,\altaffilmark{7}
X.Y.~Gao,\altaffilmark{8}
Q.X.~Geng,\altaffilmark{8}
H.W.~Guo,\altaffilmark{5}
H.H.~He,\altaffilmark{4}
M.~He,\altaffilmark{6}
K.~Hibino,\altaffilmark{9}
N.~Hotta,\altaffilmark{10}
Haibing~Hu,\altaffilmark{5}
H.B.~Hu,\altaffilmark{4}
J.~Huang,\altaffilmark{11}
Q.~Huang,\altaffilmark{7}
H.Y.~Jia,\altaffilmark{7}
F.~Kajino,\altaffilmark{12}
K.~Kasahara,\altaffilmark{13}
Y.~Katayose,\altaffilmark{3}
C.~Kato,\altaffilmark{14}
K.~Kawata,\altaffilmark{11}
Labaciren,\altaffilmark{5}
G.M.~Le,\altaffilmark{15}
J.Y.~Li,\altaffilmark{6}
H.~Lu,\altaffilmark{4}
S.L.~Lu,\altaffilmark{4}
X.R.~Meng,\altaffilmark{5}
K.~Mizutani,\altaffilmark{2}
J.~Mu,\altaffilmark{8}
K.~Munakata,\altaffilmark{14}
A.~Nagai,\altaffilmark{16}
H.~Nanjo,\altaffilmark{1}
M.~Nishizawa,\altaffilmark{17}
M.~Ohnishi,\altaffilmark{11}
I.~Ohta,\altaffilmark{10}
H.~Onuma,\altaffilmark{2}
T.~Ouchi,\altaffilmark{9}
S.~Ozawa,\altaffilmark{11}
J.R.~Ren,\altaffilmark{4}
T.~Saito,\altaffilmark{18}
T.Y.~Saito,\altaffilmark{11}
M.~Sakata,\altaffilmark{12}
T.~Sasaki,\altaffilmark{9}
M.~Shibata,\altaffilmark{3}
A.~Shiomi,\altaffilmark{11}
T.~Shirai,\altaffilmark{9}
H.~Sugimoto,\altaffilmark{19}
M.~Takita,\altaffilmark{11}
Y.H.~Tan,\altaffilmark{4}
N.~Tateyama,\altaffilmark{9}
S.~Torii,\altaffilmark{20}
H.~Tsuchiya,\altaffilmark{21}
S.~Udo,\altaffilmark{11}
H.~Wang,\altaffilmark{4}
X.~Wang,\altaffilmark{2}
Y.G.~Wang,\altaffilmark{6}
H.R.~Wu,\altaffilmark{4}
L.~Xue,\altaffilmark{6}
Y.~Yamamoto,\altaffilmark{12}
C.T.~Yan,\altaffilmark{11}
X.C.~Yang,\altaffilmark{8}
S.~Yasue,\altaffilmark{14}
Z.H.~Ye,\altaffilmark{15}
G.C.~Yu,\altaffilmark{7}
A.F.~Yuan,\altaffilmark{5}
T.~Yuda,\altaffilmark{9}
H.M.~Zhang,\altaffilmark{4}
J.L.~Zhang,\altaffilmark{4}
N.J.~Zhang,\altaffilmark{6}
X.Y.~Zhang,\altaffilmark{6}
Y.~Zhang,\altaffilmark{4}
Yi ~Zhang,\altaffilmark{4}
Zhaxisangzhu,\altaffilmark{5}
and X.X.~Zhou\altaffilmark{7} \\
(The Tibet AS$\gamma$ Collaboration)}

\altaffiltext{1}{ Department of Physics, Hirosaki University, Hirosaki 036-8561, Japan}
\altaffiltext{2}{ Department of Physics, Saitama University, Saitama
338-8570, Japan }
\altaffiltext{3}{ Faculty of Engineering, Yokohama National University, Yokohama 240-8501, Japan }
\altaffiltext{4}{ Key Laboratory of Particle Astrophysics, Institute of High Energy Physics, Chinese Academy of Sciences, Beijing 100049, China }
\altaffiltext{5}{ Department of Mathematics and Physics, Tibet University, Lhasa 850000, China }
\altaffiltext{6}{ Department of Physics, Shandong University, Jinan 250100, China }
\altaffiltext{7}{ Institute of Modern Physics, South West Jiaotong University, Chengdu 610031, China }
\altaffiltext{8}{ Department of Physics, Yunnan University, Kunming 650091, China }
\altaffiltext{9}{  Faculty of Engineering, Kanagawa University, Yokohama 221-8686, Japan}
\altaffiltext{10}{ Faculty of Education, Utsunomiya University, Utsunomiya 321-8505, Japan}
\altaffiltext{11}{ Institute for Cosmic Ray Research, University of Tokyo, Kashiwa 277-8582, Japan }
\altaffiltext{12}{ Department of Physics, Konan University, Kobe 658-8501, Japan}
\altaffiltext{13}{ Faculty of Systems Engineering, Shibaura Institute of Technology, Saitama 337-8570, Japan}

\altaffiltext{14}{ Department of Physics, Shinshu University, Matsumoto 390-8621, Japan}
\altaffiltext{15}{ Center of Space Science and Application Research, Chinese Academy of Sciences, Beijing 100080, China }
\altaffiltext{16}{ Advanced Media Network Center, Utsunomiya University, Utsunomiya 321-8585, Japan}
\altaffiltext{17}{ National Institute of Informatics, Tokyo 101-8430, Japan}
\altaffiltext{18}{ Tokyo Metropolitan College of Aeronautical Engineering, Tokyo 116-0003, Japan}
\altaffiltext{19}{ Shonan Institute of Technology, Fujisawa 251-8511,
Japan}
\altaffiltext{20}{ Advanced Research Institute for Science and
Engineering, Waseda University, Tokyo 169-8555, Japan}
\altaffiltext{21}{ RIKEN, Wako 351-0198, Japan}



\begin{abstract}
We searched  for steady PeV gamma-ray emission from the Monogem ring
 region with the Tibet air shower array from 1997 February to 2004 October
 . 
 No evidence for statistically significant gamma-ray signals was
 found in a region 111$\degr$~$\leq$ R.A.
 $<$~114$\degr$,~12$\fdg$5~$\leq$ decl. $<$ 15$\fdg$5  in the
 Monogem ring where the MAKET-ANI experiment recently claimed a positive
 detection of PeV high-energy cosmic radiation, 
although our flux sensitivity is approximately 10 times better than 
MAKET-ANI's.
 We set the most stringent integral flux upper limit at a 99$\%$
 confidence level of 4.0 $\times$
 10$^{-12}$~cm$^{-2}$~s$^{-1}$~sr$^{-1}$ above 1 
PeV on diffuse gamma rays extended in the 3$^{\circ}$ $\times$ 3$^{\circ}$ region.
\end{abstract}


%% subject header.  Objects should be in the appropriate "individual"
%% headers (e.g. quasars: individual, stars: individual, etc.) with the
%% additional provision that the total number of headers, including each
%% individual object, not exceed six.  The \objectname{} macro, and its
%% alias \object{}, is used to mark each object.  The macro takes the object
%% name as its primary argument.  This name will appear in the paper
%% and serve as the link's anchor in the electronic edition if the name
%% is recognized by the data centers.  The macro also takes an optional
%% argument in parentheses in cases where the data center identification
%% differs from what is to be printed in the paper.

\keywords{ (stars:) supernovae: individual (Monogem Ring) --- gamma rays : observations}


\section{Introduction}

In a recent observation at PeV energies, the MAKET-ANI air shower
experiment at Mount Aragats
(E44$\degr$10$\arcmin$, N40$\degr$30$\arcmin$N; 3200 m above sea
level) claimed a detection of significant excess (6$\sigma$) of
cosmic-ray events within a 3$\degr$~$\times$~3$\degr$ search 
window~(111$\degr$~$\leq$ $\alpha$ $<$ 114$\degr$,
12$\fdg$5 $\leq$ $\delta$ $<$ $15\fdg5$) in the
 Monogem ring region using the air shower data recorded from 1997 to
2003\\ \cite{Chilingarian2003}
Naturally, the significant excess may be attributed to PeV gamma
rays, because the Larmor~radius $R_{\rm L}$ $\sim$ 0.4 $/$ $Z$~pc
at 10$^{15}$~eV in the galactic magnetic field of 3 $\mu$G is too
small to reach the Earth without deflection compared with the distance of
300~pc between the Earth and the Monogem~ring, and the mean decay
length of a neutron $\lambda$ $\sim$ 10~pc at 10$^{15}$~eV is also too short.

The Monogem~ring is a diffuse (extended with a diameter of 25$\degr$ in
the sky) supernova remnant (SNR) associated with the radio pulsar
PSR~B0656+14 at a distance of approximately 300~pc
\cite{Thorsett2003}.  This SNR is a bright source in the soft X-ray
region. According to a detailed observation of the Monogem~ring made
by the $ROSAT$ X-ray survey \cite{Plucinsky1996}, the average
temperature of thermal emission is 6.15 in log($T$/1 K).  If the
Monogem~ring is modeled as an SNR in the adiabatic stage of evolution,
the initial ambient density is estimated to be
5.2~$\times$~10$^{-3}$~cm$^{-3}$, the initial explosion energy is estimated to
 be 1.9~$\times$~10$^{50}$~erg, the radius to the shock front is estimated to 
be 66.5~pc, and the age is estimated to be 86 thousand years. 
The Monogem ring has been considered to be a possible
candidate for the particle acceleration site of cosmic electrons \cite{Kobayashi2004}
and nuclei \cite{Thorsett2003} in our Galaxy.

The cosmic-ray flux we observe at the Earth roughly shows a power-law
spectrum at energies from 10$^{9}$ to 10$^{20}$~eV.  There is a
general consensus that the stochastic shock acceleration at SNRs in our
Galaxy could explain the cosmic-ray energy spectrum up to approximately
$10^{15}$~eV.  The maximum energy is estimated to be only $\sim Z \times
10^{14}$ eV by parallel shock acceleration where the normal axis of shock
front is parallel to the direction of the interstellar magnetic field.
On the other hand, the oblique shocks accelerate particles more
efficiently than the parallel shocks \cite{Jokipii1987}. The maximum
energy based on the oblique shock acceleration increases by several
orders of magnitude, and could explain the cosmic-ray energy spectrum up to
approximately $10^{17}$~eV \cite{Kobayakawa2002}.  In this case, the
cosmic rays beyond $10^{17}$~eV are assumed to be of extragalactic
sources, such as active galactic nuclei and gamma-ray burst
.  Among recent models, the ``single source (SS) model''
\citep{Berezhko1996,Erlykin} is interesting in that a single nearby SNR,
 such as the Monogem~ring, mainly contributes to the cosmic-ray 
intensity observed at
the Earth.  To explain the shape and intensity of the cosmic-ray energy
spectrum using the SS model, the most likely parameters of the single SNR
should be 300--350 pc distant and 90--100 thousand years old
\cite{Erlykin2003}.  If such a strong cosmic-ray accelerator lies near
the Earth, we may observe an anisotropy of cosmic rays from its
direction, or we may detect high-energy photons that are emitted from these
high-energy charged particles by the nonthermal processes.

In this paper, we report on the search for
diffuse/pointlike PeV gamma-ray emission based on the data recorded
from 1997 to 2004 around the Monogem~ring region by a large air shower
array with a total area of 36,900~m${^2}$ constructed in Tibet.

\section{Experiment}

The Tibet air shower experiment has been successfully operated at
 Yangbajing (E90$^{\circ}$31$\arcmin$, N30$^{\circ}$06$\arcmin$; 4300
 m above sea level) in Tibet, China since 1990.  The Tibet~I array was
 constructed in 1990 \cite{Tibet1} and it was gradually expanded by 1994 to the
 Tibet~II that consisted of 185 fast-timing (FT) scintillation
 counters placed on a 15 m square grid covering 36,900~m$^2$ and 36
 density (D) scintillation counters around the FT-counter array.  Each
 counter has a plastic scintillator plate (BICRON BC-408A) of 0.5~m$^2$
 in area and 3~cm in thickness. A 0.5~cm thick lead
 plate is put on the top of each counter in order to increase the
 counter sensitivity by converting gamma rays into electron-positron
 pairs in an electromagnetic shower \citep{Lead,Amenomori1990}.
 All the FT counters are equipped with a
 fast-timing photomultiplier tube (FT-PMT; Hamamastu H1161) measuring up
 to 15 particles, and 52 out
 of 185 FT counters, which are arrayed at 30 m lattice intervals, are
 also equipped with a wide dynamic range PMT (D-PMT; Hamamatsu
 H3178) measuring up to 500 particles. The time and charge information of
 the FT-PMTs is recorded, while only the charge information of D-PMT is recorded.
 All the D counters are also equipped with FT-PMT and D-PMT, where only
 charge information of both PMTs is recorded. An event trigger signal is
 issued when any fourfold coincidence occurs in the FT counters
 recording more than 0.6 particles.
 The mode of 
 energy of the triggered events in Tibet~II is $\sim$10 TeV. 

 From 1996 to 2003, we upgraded the array, and, at present, it consists of 
 761 FT counters covering 50,400~m$^2$ and 28 D counters around
 them. In the inner 36,900 m$^2$, FT counters are deployed at 7.5 m lattice
 intervals. Since 1999 October, we have called this upgraded array Tibet III. 
The mode of energy of the triggered events in
 Tibet III is 3 TeV.  Using Tibet III, we observed the multi-TeV
 gamma-ray flares from Mrk~421 in 2000 and 2001 \cite{mrk421} and the
 cosmic-ray anisotropy \citep {aniso,side}, surveyed the
 northern TeV gamma-ray sky \cite{all}, and measured the all-particle
 energy spectrum \cite{ozawa} and chemical composition of the cosmic rays in the knee region \cite {compo}.

\section{Analysis}

In the present paper, we employ the data obtained by the 185 FT counters
  and the 36 D counters corresponding to the Tibet II array
  configuration for the whole period in order to simplify the
  analysis. 
  We collected 1.6~$\times$~10$^8$ air shower events during 1717
  detector live
  days from 1997 February 15 to 2004 October 10 after the quality
  cut and the event selection based on the following simple criteria:\\
\\
1. $Air shower core location.$---Among the three hottest counters in each
  event, two should be contained in the inner 36,900 m$^2$.\\
2. $Shower size.$---$\Sigma \rho _{\rm D}$ which is the sum of the number 
of particles per m$^2$ counted by the 36 D counters and 52 out of 185 FT counters that have a D-PMT, should be more than 100.\\
3. $Zenith angle.$--- The zenith angle of the arrival direction should be less than 40$^{\circ}$.
%(4) Mean residual error : The mean residual error should be less than
%1~m, where the mean residual error is the mean distance between the
%reconstructed shower front and the each particle position at the time
%when it hits a counter.

To examine the performance of the Tibet II array, we use a Monte Carlo (MC) simulation. We employ the CORSIKA version 6.200 code \cite{heck} for the generation of air shower events and the EPICS uv8.00
code (K.Kasahara 2005)\footnote{Additional information on Kasahara (2005)
is available at
http://eweb.n.kanagawa-u.ac.jp/\~{}kasahara/ResearchHome/EPICSHome/.}
for the detection of shower particles with scintillation counters,
respectively. The primary gamma rays are sampled from the Monogem ring
orbit assuming a differential power-law spectrum with spectral index
$-$2.0 from 10 TeV to 30 PeV.

According to the MC simulation including the quality cut and the event selection,
the mode energy of gamma rays is 150~TeV, the angular resolution is less
than 0$\fdg$3, and the effective area for gamma rays is nearly 2.5 $\times$ 10$^4$ m$^2$ for a 3$^{\circ}$ $\times$
3$^{\circ}$ search window in a diffuse source analysis and 1.6 $\times$ 10$^4$ m$^2$ for a 0$\fdg$5
$\times$ 0$\fdg$5 search window in a pointlike source analysis. The gamma-ray energy is
estimated from $\Sigma \rho_{\rm D}$ by the MC simulation and
the energy resolution is less than 30$\%$ above 150 TeV. The systematic
pointing error is estimated to be less than 0$\fdg$02 by the Moon's
shadow in cosmic rays \cite{mrk421}. The source positional uncertainty
is typically less than 0$\fdg$1 at 150 TeV  for gamma rays, assuming a 5
$\sigma$ significance level.
The center of the Monogem ring
($\delta=14\degr$) stays
in the field of view (with a zenith angle of $<$40$\degr$) for 380 days out
of 1717 live days.

 Subsequently, we use the right ascension scan method to search for PeV gamma-ray sources that follows the same analysis method and parameters employed by the MAKET-ANI experiment \cite{Chilingarian2003}.
First, each event is sorted by its arrival right ascension and
declination into a $\Delta\alpha$ $\times$ $\Delta\delta$ = 3$^{\circ}$
$\times$ 3$^{\circ}$ rectangular cell. Off-source events are
taken from all the cells (except the on-source cell) in the same declination
band as the on-source cell. The significance of the
source in each cell is calculated based on exsquation (17) of Li $\&$
Ma \cite{lima}.

 We scanned the celestial sky  in the declination band from
 $-5\fdg$5 to  66$\fdg$5 in the whole right ascension range 0$^{\circ}$
 -- 360$^{\circ}$. 
 We also scanned the whole Monogem ring region and the region around it 
in the declination band from $0\degr$ to $30\degr$  in the right ascension range 80$^{\circ}$
 -- 130$^{\circ}$ with a $\Delta\alpha$ $\times$ $\Delta\delta$ = 0$\fdg$5
 $\times$ 0$\fdg$5 search window analysis for a pointlike source.

\section{Results}
\begin{figure}

\plotone{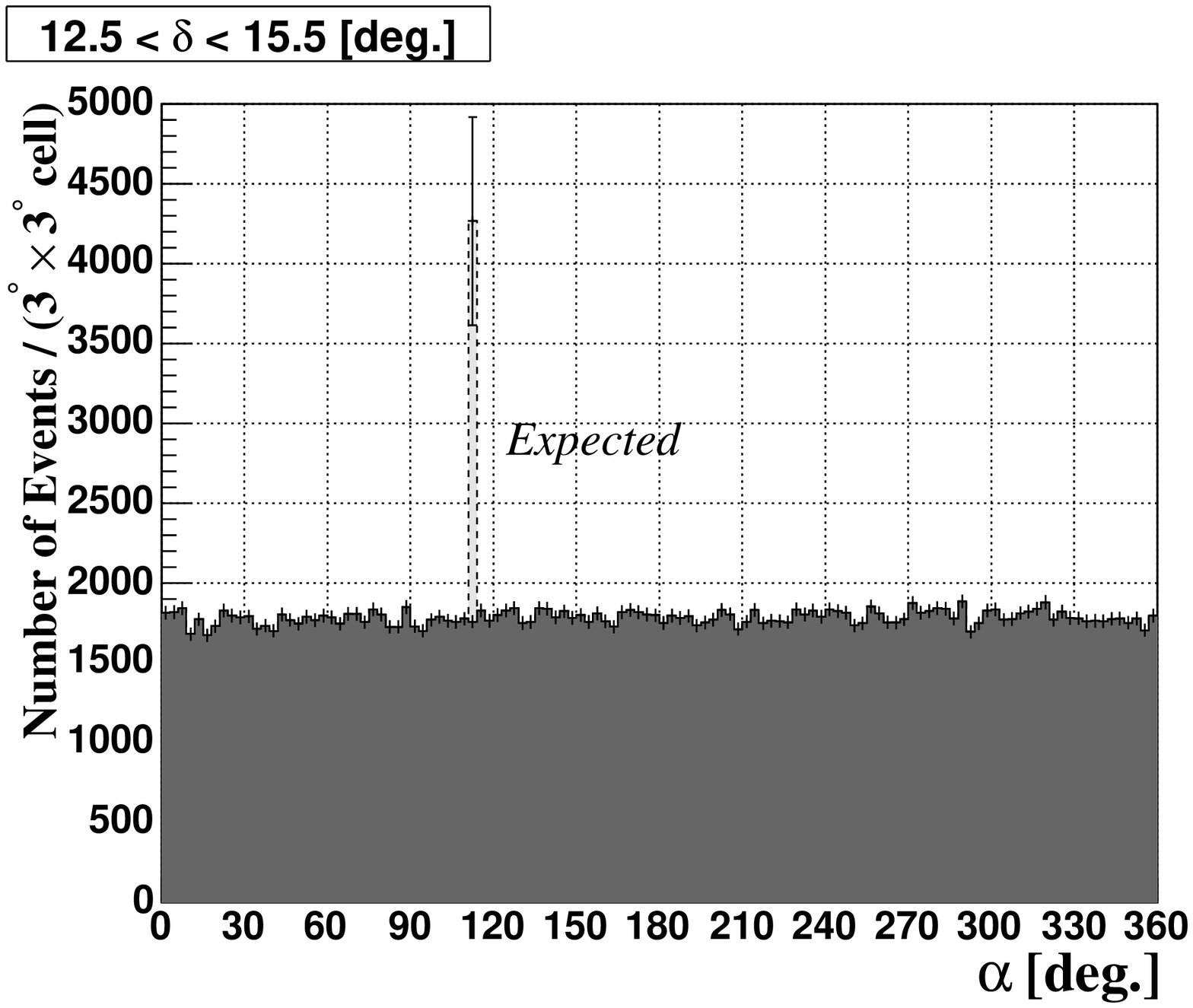}
\caption{Number of events above 1~PeV in each of the 120
 cells
 in the declination band of 12$\fdg$5--15$\fdg$5 observed by the Tibet
 air shower array during 1717 live days from 1997 February~15 to 2004 October~10. The shaded histogram represents our actual result, while the dashed histogram denotes the number of events expected from the MAKET-ANI result \cite{Chilingarian2003}.} \label{onebarrel}
\end{figure}

\begin{figure}
\plotone{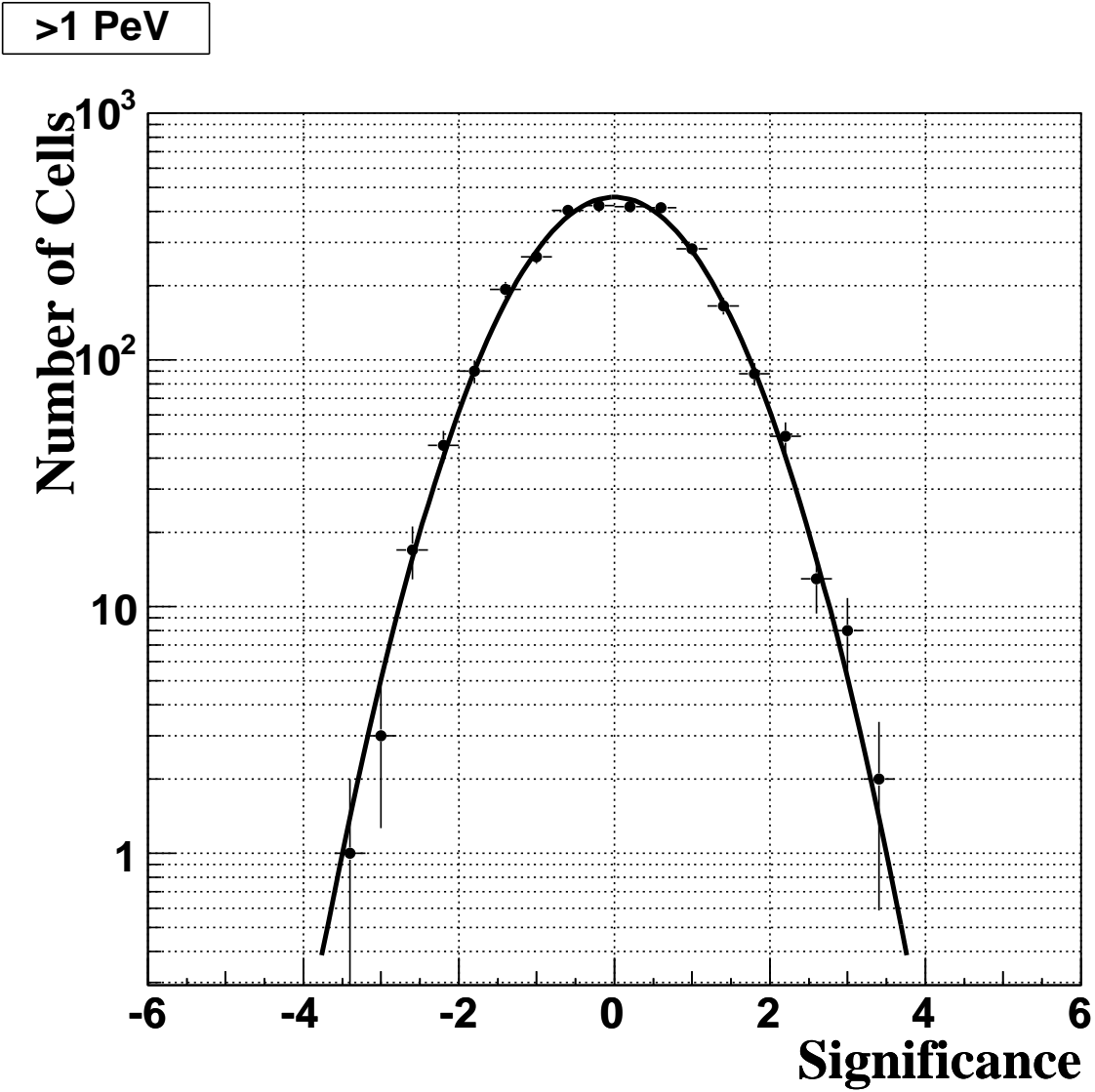}
\caption{Significance distribution of 2880 $\Delta\alpha \times
 \Delta\delta$ = 3$^{\circ}$$\times$3$^{\circ}$ cells in the declination band
 --5$\fdg$5~$\leq$~$\delta$~$<$~66$\fdg$5 above 1~PeV. Each cell is
 independent of one another. The solid line indicates a normal
 Gaussian fit to the data.} \label{twobarrel}
\end{figure}

Figure~\ref{onebarrel} shows the number of events in each of the 120 cells in the declination band 
12${\fdg}$5--15${\fdg}$5 by the $\Delta\alpha$ $\times$ $\Delta\delta$ =
3$^{\circ}$ $\times$ 3$^{\circ}$ search window analysis with $ \Sigma \rho _{\rm D} > 1000 $ (corresponding to $>1$~PeV). The shaded histogram denotes our actual result, and the dashed one is the expected excess from the MAKET-ANI result.
The MAKET-ANI experiment detected a significant excess in the direction 111$\degr$ $\leq$ $\alpha$ $<$ 114$\degr$, 12$\fdg$5 $\leq$ $\delta$ $<$ 15$\fdg$5 at the 6 $\sigma$ statistical significance. 
But no significant signal was detected by the Tibet air shower
array($-0.6 \sigma$).
It should be noted that the Tibet air shower array accumulated about 100 times as much statistics as the MAKET-ANI experiment, corresponding to an approximately 10 times better sensitivity, which is easily calculated by the number of background events as summarized in Table~1.
In addition, the significance distribution as shown in Figure~\ref{twobarrel} implies
the absence of such a bright and diffuse source above 1~PeV, even in the all sky survey ($0\degr \leq \alpha < 360\degr$, $-5\fdg5 \leq \delta < 66\fdg5$).\\
  Taking the pointing errors of both experiments into account, we scanned the area of $ \alpha$~$\pm$~5$^{\circ}$, $\delta$~$\pm$~5$^{\circ}$ around the source direction claimed by the MAKET-ANI experiment by sliding the center of the 3$^{\circ}$ $\times$ 3$^{\circ}$ search window by 0$\fdg$5 steps in right ascension and declination; however, no significant excess was found again. 

 Table~1 demonstrates the energy dependence of the result in the region
 suggested by the MAKET-ANI experiment by the 3$^{\circ}\times
 3^{\circ}$ search window analysis, also confirming no signal
 detection. As the MAKET-ANI experiment detected a significant excess in
 the various energy thresholds of $>$800~TeV, $>$1~PeV, and $>$2~PeV, we
 would have detected a significant signal at $50\pm10$ $\sigma$
 (estimated from the number of on-source and off-source events based on
equation (17) of Li $\&$ Ma \cite{lima}, which is a different definition from
 MAKET-ANI's) in at least one energy threshold even if the relative
 energy scale uncertainty between the two experiments differed by a
 factor of 2.

\begin{figure}
\plotone{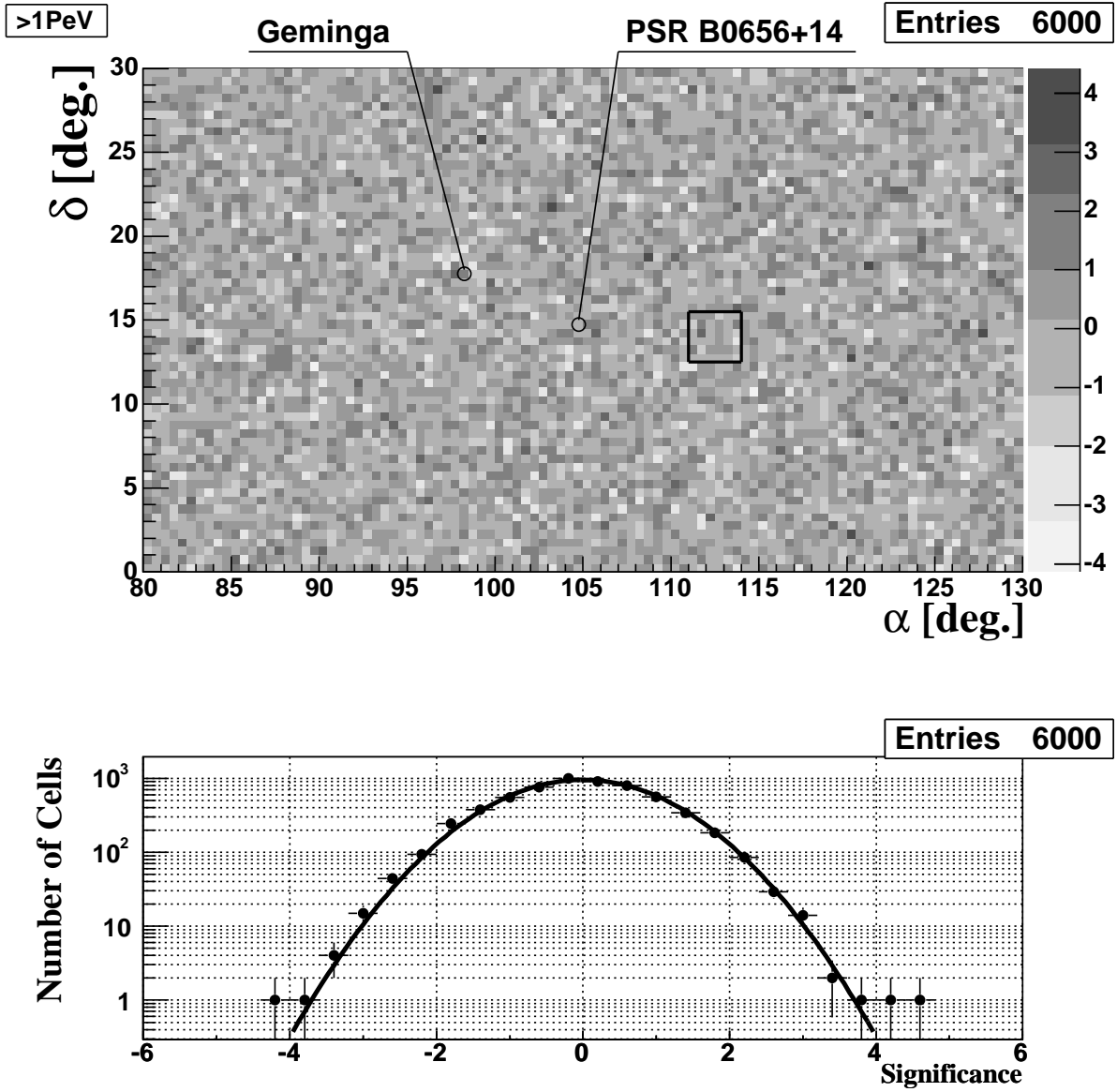} \caption{$Top$: Significance map in the Monogem
 Ring region and the region around it above 1~PeV. Each cell is independent of one another. PSR
 B0656+14 is located at the center of the region. The rectangular region indicates the area
where the MAKET-ANI experiment claimed a positive detection of a signal.
 $Bottom$: Significance distribution in the region shown in the map. The
 solid line indicates a normal Gaussian fit to the data. The two cells
with $>4$ $\sigma$ are located at ($\alpha$, $\delta$) =
(108$\fdg$75,~18$\fdg$75) and (120$\fdg$25,~12$\fdg$75), respectively.}
\label{threebarrel}
\end{figure}

Figure~\ref{threebarrel} shows the significance map of the whole Monogem ring region and the region 
 around it (80$^{\circ}$~$\leq$~$\alpha$~$<$~130$^{\circ}$,
 0$^{\circ}$~$\leq$~$\delta$~$<$~30$^{\circ}$) based on a finer window
 search of 0$\fdg$5 $\times$ 0$\fdg$5 cells at energies $>1$~PeV. Again,
 no significant signal was found. There are two directions with a
 significance $>4\sigma$ at~($\alpha$,~$\delta$) = (108$\fdg$75,~18$\fdg$75), (120$\fdg$25,~12$\fdg$75) that are very far from the point where the MAKET-ANI experiment claimed the detection of a signal.
 The expected number of directions above 4$\sigma$ in a normal Gaussian distribution with 6000 trials is 0.19, and  the probability of getting more than 2 in a Poisson distribution with mean value of 0.19 is 1.6 $\times$ 10$^{-2}$. Therefore, the deviation may be due to statistical fluctuations. 
The results of the 0$\fdg$5 $\times$ 0$\fdg$5 window searches did not
 show any significant energy dependence, and no significant deviation of
 significance distribution from a normal Gaussian distribution was found at
 the other energy thresholds ($>500$~TeV, $>800$~TeV, and $>2$~PeV).

\section{Discussions}

No evidence for statistically significant gamma-ray signals was found in
a region 111$\degr~\leq \alpha<~114\degr$,~12$\fdg5~\leq \delta < 15\fdg$5  in the Monogem Ring where the MAKET-ANI experiment recently claimed a positive
 detection of PeV high energy cosmic radiation, although our flux
 sensitivity is approximately 10 times better than MAKET-ANI's.
We set the most stringent integral flux upper limit at a 99$\%$
confidence level of 1.1 $\times$ 10$^{-14}$ cm$^{-2}$s$^{-1}$/(2.66 $\times$ 10$^{-3}$sr) =~4.0 $\times$ 10$^{-12}$~cm$^{-2}$~s$^{-1}$~sr$^{-1}$ on steady diffuse gamma rays $>1$~PeV extended within the rectangular region in the Monogem Ring assuming a differential spectral index $-2.0$.
One of the potential possibilities for explaining the discrepancy between the 
two experiments could be transient emission that occurred at occasions when we stopped data
acquisition for annual maintenance, calibration, upgrading jobs, etc.
Another could be the strong transient emission of PeV gamma rays that
incidentally or periodically occurred during the 3 hr when only
the MAKET-ANI experiment could observe because the two experimental
sites are separated by about 45$\degr$ in longitude. Third, the excess
events that they detected might be due to statistical fluctuation
\citep{ChilingarianICRC}.

  No significant signal was found in the whole Monogem Ring region based
  on a 0$\fdg$5 $\times$ 0$\fdg$5 window search for a pointlike source
  at energies $>1$ PeV.  We also set 99$\%$ confidence-level flux upper
  limits of 2.6 and 5.4 $\times$
  10$^{-15}$ cm$^{-2}$ s$^{-1}$ on the steady gamma rays $>1$~PeV from
  PSR~B0656+14 and Geminga, respectively, assuming pointlike sources
  with a differential energy spectral index $-$2.0. 
  The KASCADE group also reported that no significant signal was seen in
  the sub-PeV region at
  the suggested location by the MAKET-ANI experiment and PSR~B0656+14
 by a pointlike source analysis \cite{Antoni2004}.
  Furthermore, we reported on the result of wide sky survey for steady TeV
  gamma-ray pointlike sources elsewhere \cite{all}, although no
  significant point source was found in the Monogem Ring region above a
  few TeV and above 10 TeV. 

 \acknowledgments

 This work is supported in part by Grants-in-Aid for Scientific
 Research on Priority Area (712) (MEXT) and also for Scientific Research (JSPS)
  in Japan,  and by the Committee of the Natural Science Foundation and by the
 Chinese Academy of Sciences in China. 




% TABLE3.TEX -- Sample table 3.

% In this example we could not use the \phs command in place of the 
% \phantom{-} command since \phs is already in math mode.  We could of
% course have reformatted the table, but chose not to do so.

\begin{deluxetable}{lrrrr}
\tablecolumns{5}
\tablewidth{40pc}

\tablecaption{Enegy dependence of the number of events in the cell
 111$^{\circ}$~$\leq$~$\alpha$~$<$~114$^{\circ}$,
 12$\fdg$5~$\leq$~$\delta\leq$~15$\fdg$5 ($N^{\mathrm{ON}}$) and
 the background ($N^{\mathrm{BG}}$)}

\tablehead{
\colhead{Primary Gamma-ray} & \colhead{     } & \colhead{         } & \colhead{           } & \colhead{          } \\
\colhead{Energy\tablenotemark{a}  [TeV]}&
\colhead{\hspace{1cm}$N^{\mathrm {BG}}_{\mathrm {Tibet}}$\tablenotemark{b} } &
\colhead{\hspace{1cm}$N^{ \mathrm {ON}}_{ \mathrm {Tibet}}$ } & 
\colhead{\hspace{1cm}$N^{ \mathrm {BG}}_{ \mathrm {MAKET}}$\tablenotemark{b} }    & 
\colhead{\hspace{1cm}$N^{ \mathrm {ON}}_{ \mathrm {MAKET}}$ }
}
\startdata
$>$500 &   5664.7 &  5605  &  58 & 84 \\ 
$>$800 &   2609.6 &  2618  &  26 & 57 \\ 
$>$1000 ($N_e$ $>$ 10$^6$)&  1785.9 &  1759 &  18 & 43 \\ 
$>$2000 &  506.3 & 490   &  4 & 13 \\
\enddata

\tablenotetext{a}{The primary gamma-ray energy of the MAKET-ANI
 experiment was estimated by the air shower size reported in
 \cite{Chilingarian2003}, where $N_e$ $>$10$^6$ at MAKET-ANI altitude
 corresponds to $>$1~PeV \cite{Erlykin} and where we scaled the other
 energies, while that of the Tibet air shower array is estimated by the
 air shower size $\Sigma \rho  _{\rm D}$, described in the text, in the Tibet air shower array.}

\tablenotetext{b}{Mean value of the background cells in the declination band
 12$\fdg$5~$\leq$~$\delta\leq$~15$\fdg$5. Subscripts ``Tibet'' and
 ``MAKET'' represent our result and the MAKET-ANI result, respectively. }

%\tablenotetext{b}{The mean value of all the 120 cells in the declination band 12$\fdg$5 $\leq$~$\delta$~$<$~15$\fdg$5 observed by the Tibet air shower array and the MAKET-ANI experiment.}

%\tablenotetext{c}{The number of events in the cell 111$^{\circ}$~$\leq$~$\alpha$~$<$~114$^{\circ}$, 12$\fdg$5~$\leq$~$\delta$~15$\fdg$5 observed by the Tibet air shower array and the MAKET-ANI experiment.}

\end{deluxetable}

\end{document}